\documentclass[twoside]{article}
\usepackage{qic,epsfig}

\newcommand{\be}{\begin{equation}}
\newcommand{\ee}{\end{equation}}
\newcommand{\bea}{\begin{eqnarray}}
\newcommand{\eea}{\end{eqnarray}}
\newcommand{\RR}{\rangle}
\newcommand{\LL}{\langle}
\newcommand{\RL}{\RR\LL}

\begin{document}
\setlength{\textheight}{8.0truein}    

\runninghead{An observable measure of  entanglement for pure states
of multi-qubit systems} 
            {G.K. Brennen}

\normalsize\textlineskip
\thispagestyle{empty}
\setcounter{page}{619}


\vspace*{0.88truein}

\alphfootnote

\fpage{619}

\centerline{\bf AN OBSERVABLE MEASURE OF ENTANGLEMENT FOR PURE}
\vspace*{0.035truein}
\centerline{\bf STATES OF MULTI-QUBIT SYSTEMS}
\vspace*{0.37truein}
\centerline{\footnotesize
GAVIN K. BRENNEN}
\vspace*{0.015truein}
\centerline{\footnotesize\it Quantum Processes Group, National Institute of Standards and Technology}
\baselineskip=10pt
\centerline{\footnotesize\it Gaithersburg, MD  20899-8423, USA}
\vspace*{0.225truein}

\vspace*{0.21truein}

\abstracts{Recently, Meyer and Wallach [D.A.~Meyer and N.R.~Wallach (2002), J. of Math. Phys., 43, pp. 4273] proposed a measure of multi-qubit entanglement that is a function on pure states.  We find that this function can be interpreted as a physical quantity related to the average purity of the constituent qubits and show how it can be observed in an efficient manner without the need for full quantum state tomography.  A possible realization is described for measuring the entanglement of a chain of atomic qubits trapped in a 3D optical lattice.  
}{}{}

\vspace*{10pt}

\vspace*{3pt}

\vspace*{1pt}\textlineskip    
\section{Introduction}        
Characterizing many body entanglement has become an active area of research both for understanding dynamics of interacting systems and identifying resources for quantum information processing (QIP).  Many tasks in QIP involve mapping a configuration of spins prepared in a product state to a number of different classes of entangled states.  In order to characterize the dynamics of entanglement as well as the performance of the mapping, it would be helpful to have a single parameter that quantifies the amount of multi-particle entanglement contained in the system at any given time step.  A good measure of entanglement would be one that faithfully captures the nature of nonlocal quantum correlations in a quantum state and as such should be a mapping from states to reals that is non increasing under local operations and classical communication on average.  Such measures are known as entanglement monotones \cite{Vidal}.  There is no unique function that satisfies these properties, and for multi-particle systems the problem is compounded by the fact that different degrees of entanglement can be shared between different partitions of the total system.  Several important results have been found.  For example, for bipartite pure states, a family of entanglement monotones is given by the Schmidt coefficients of the quantum state \cite{Vidal}.  For more than two subsystems, various functions have been introduced to quantify entanglement.  These include the fully entangled fraction \cite{Grondalski}, which measures the ability of a state to perform tasks such as teleportation and dense coding, polynomial invariants, which involve moments of the reduced state eigenvalues \cite{Barnum}, and the Schmidt measure \cite{Schmidt}, which is related to the minimum number of terms in the product state expansion of a state.  The later is an entanglement monotone, as are some polynomial invariants.

\section{Pure state entanglement}
\noindent

A difficulty with many entanglement measures is that they do not directly correspond to the outcome of a set of observables on the state.  It has been shown that with a small number of identical copies of a state, one can measure quantum entanglement between two systems without prior knowledge of the state \cite{Horodecki}.  For more than two systems, most entanglement measures require knowledge of the state itself, which involves performing quantum state tomography.  For a state residing in a $d$ dimensional Hilbert space, this means a minimum of $d^{2}$ measurements for each statistical sample must be made on the system.  It would be advantageous, especially with regard to long spin chains, to have an easily measured quantity of multi-qubit entanglement.  
Meyer and Wallach \cite{Meyer} have proposed a multi-qubit entanglement measure which was derived as a formal function on many qubit states but has a simple physical interpretation.  They introduce a function on pure states of $n$ qubits defined as,
\be
Q(|\psi\RR)=\frac{4}{n}\sum_{k=1}^{n}D(|\tilde{u}^{k}\RR,|\tilde{v}^{k}\RR),\;
\ee
where $|\tilde{u}^{k}\RR$ and $|\tilde{v}^{k}\RR$ are vectors in ${\bf C}^{2n-2}$ which are non-normalized (indicated by the $\tilde{}\ $) and obtained by projecting on state $|\psi\RR$ with local projectors on the $k$th qubit,
\be
|\psi\RR=|0\RR_{k}\otimes|\tilde{u}^{k}\RR+|1\RR_{k}\otimes|\tilde{v}^{k}\RR.\;
\ee
The function $D(|\tilde{u}^{k}\RR,|\tilde{v}^{k}\RR)$ measures a ``distance" between the two vectors $|\tilde{u}^{k}\RR$ and $|\tilde{v}^{k}\RR$.  It is obtained by taking the generalized cross product:
\be
D(|\tilde{u}^{k}\RR,|\tilde{v}^{k}\RR)=\sum_{i<j}|\tilde{u}^{k}_{i}\tilde{v}^{k}_{j}-\tilde{u}^{k}_{j}\tilde{v}^{k}_{i}|^2.\;
\ee
The quantity $Q(|\psi\RR)$ is invariant under local unitaries because the effect of applying $U_{j}$ is to multiply each term $D(|\tilde{u}^{k}\RR,|\tilde{v}^{k}\RR)$ by $|det(U_{j})|^{2}=1$.  One could also write the state $|\psi\RR$ in the Schmidt decomposition over the bipartite division of qubit $k$ and the other qubits as:
\be
|\psi\RR=|\bar{0}\RR_{k}\otimes|\tilde{x}^{k}\RR+|\bar{1}\RR_{k}\otimes|\tilde{y}^{k}\RR,\;
\ee
where $\LL\tilde{x}^{k}|\tilde{y}^{k}\RR=0$, and $\{|\bar{0}\RR_{k},|\bar{1}\RR_{k}\}$ are related to $\{|0\RR_{k},|1\RR_{k}\}$ by a local unitary operator $U_{k}$.  The purity of the state of qubit $k$ is therefore $Tr[\rho_{k}^{2}]=\LL\tilde{x}^{k}|\tilde{x}^{k}\RR^{2}+\LL\tilde{y}^{k}|\tilde{y}^{k}\RR^{2}$.  By invariance of the generalized cross product under local unitaries, $D(|\tilde{u}^{k}\RR,|\tilde{v}^{k}\RR)=D(|\tilde{x}^{k}\RR,|\tilde{y}^{k}\RR)$ and this quantity can be related to the norm of an antisymmetric tensor $M_{k}=|\tilde{x}^{k}\RR\LL\tilde{y}^{* k}|-|\tilde{y}^{k}\RR\LL\tilde{x}^{* k}|$ as \cite{Manny} 
\be
\begin{array}{lll}
D(|\tilde{x}^{k}\RR,|\tilde{y}^{k}\RR)&=&\sum_{i<j}{|\tilde{u}^{k}_{i}\tilde{v}^{k}_{j}-\tilde{u}^{k}_{j}\tilde{v}^{k}_{i}|^2}\\
&=&\frac{1}{2}\sum_{i,j}(M_{k}^{\dagger})_{ij}(M_{k})_{ji}\\
&=&\frac{1}{2}Tr[M_{k}^{\dagger}M_{k}]\\
&=&\LL \tilde{x}^{k}|\tilde{x}^{k}\RR\LL \tilde{y}^{k}|\tilde{y}^{k}\RR\\
&=&1/2 (1-Tr[\rho_{k}^{2}]).\;
\end{array}
\ee
Therefore,
\be
Q(|\psi\RR)=2(1-1/n\sum_{k=0}^{n-1}Tr[\rho_{k}^{2}]).\;
\label{Q}
\ee
This expression elucidates the physical meaning of the multi-particle entanglement as an average over the entanglements of each qubit with the rest of the system.  Put in this form, the entanglement measure can also be identified with the purity of a state relative to the Lie algebra of local observables for $n$ qubits within the generalized-entanglement framework of Barnum {\it et al.} \cite{genent}.

The measure $Q$ is linearly related to the average purity of each qubit and it is easy to check that it satisfies the following properties:
(i)  $0\leq Q(|\psi\RR)\leq 1$.  $Q(|\psi\RR)=0$ iff $|\psi\RR$ is a product state and $Q(|\psi\RR)=1$ for some entangled states,
(ii) $Q(|\psi\RR)$ is invariant under local unitaries $U_{j}$.

Property (i) follows from the fact that any reduced state, $\rho_{j}$, is at most rank two and because its eigenvalues sum to one, $1/2\leq Tr[\rho_{j}^{2}]\leq 1$.  When $|\psi\RR$ is a product state, $Tr[\rho_{j}^{2}]=1 \ \forall j$ implying $Q(|\psi\RR)=0$.  Conversely, $Q(|\psi\RR)=0$ implies $\sum_{j=0}^{n-1}Tr[\rho_{j}^{2}]=n$ but because $Tr[\rho_{j}^{2}]$ is bounded above by 1 this can only be true if $Tr[\rho_{j}^{2}]=1\ \forall j$.  $Q(|\psi\RR)=1$ only if $Tr[\rho_{j}^{2}]=1/2 \ \forall j$ meaning the reduced state of every qubit is maximally mixed.  Property (ii) is a consequence of the fact that the eigenvalues of the reduced states are invariant under local unitaries.

Many pure states obtain the maximum value of $Q$ including the $n$-partite GHZ state:  $1/\sqrt{2}(|0\ldots0\RR+|1\ldots1\RR)$, and the ``cluster state" \cite{Briegel} defined as the $n$ qubit state:
\be
|\Psi_{n\ clus}\RR=\frac{1}{2^{n/2}}\bigotimes_{a=0}^{n-1}(|0\RR_{a}\sigma_{z}^{a+1}+|1\RR_{a}),\;
\ee
with the convention $\sigma_{z}^{n}={\bf 1}$.  These two states are inequivalent under local unitaries as best seen by comparing the Schmidt number of each, defined as the minimum number of product states over all subsystems necessary to describe the state.  In the former it is two, while for the later it is $2^{\lfloor n/2\rfloor}$.  It is relatively easy to calculate the entanglement of certain pure states.  For instance, the $n$ qubit $W$-state \cite{dur} defined as $|W_{n}\RR=\frac{1}{\sqrt{n}}\sum_{\pi}|\pi(0\dots01)\RR$, where the sum is over the $n$ cyclic permutations, has an entanglement value $Q(|W_{n}\RR)=4 (n-1)/n^{2}$.   Meyer and Wallach \cite{Meyer} calculate the entanglement of several other states such as the eigenstates of the Heisenberg interaction and single logical qubit states encoded in the five qubit error correction code.  

A deficiency of $Q$ as an entanglement measure is that it cannot distinguish sub-global entanglement.  For example, the two states $|\Phi\RR=1/\sqrt{2}(|00\RR_{1,2}+|11\RR_{1,2})\otimes 1/\sqrt{2}(|00\RR_{3,4}+|11\RR_{3,4})$ and $|\Psi\RR=1/\sqrt{2}(|0000\RR_{1,2,3,4}+|1111\RR_{1,2,3,4})$ both have $Q$ values equal to 1.  One could distinguish different states with the same $Q$ value by comparing the Schmidt numbers over various bipartite divisions.  For instance, the Schmidt number of $|\Phi\RR$ over the division into two sets of qubits \{1,2\} and \{3,4\} is 1, while for $|\Psi\RR$ it is 2.

\subsection{Observation of entangled states}
A benefit of the reformulation of measure $Q$ in Eq.~\ref{Q} is that it identifies a relatively small number of parameters that need to be estimated in order the measure the multi-qubit entanglement.  The computation of an arbitrary function on a state of $n$ qubits requires good estimates of all $2^{2n}$ parameters of the state.  By the formulation of $Q$ in terms of the average purity of the qubits, it is only necessary to estimate a number of parameters linear in $n$, in particular only the purity of each qubit.

The purity of a single quantum state can be observed using an entanglement witness as proposed  by Filip \cite{Filip}.  He shows that given two systems, A and B, initially in the product state $\rho\otimes\rho$, one can measure a quantity proportional to $Tr[\rho^{2}]$ by interacting the two systems with a single ancillary qubit and measuring the ancilla. The ancilla is prepared in the superposition $1/\sqrt{2}(|0\RR+|1\RR)$ and plays the role of the control qubit in a controlled SWAP (c-SWAP) gate acting on the two systems.  When the ancilla is measured in the $\sigma_{x}$ basis, the probability to measure population in the $\pm$ eigenstate is:  $p(\pm)=1/2(1\
\pm Tr[\rho^{2}])$.  In \cite{Filip} it is claimed that this constitutes a quantum non-demolition (QND) measurement of purity of the quantum state $\rho$ because the output of the quantum process is a balanced mixture of the two quantum states regardless of the measurement outcome.  This, in fact, is not the case because the systems generally become entangled with each other.  The measurement process described above constitutes a measurement of the SWAP operator which projects the two systems into the symmetric or antisymmetric subspaces of the input state $\rho\otimes\rho$.  Specifically, if we write the state $\rho$ in an eigendecomposition:  $\rho=\sum_{i}p_{i}|\phi_{i}\RL\phi_{i}|$, then a selective measurement of the ancilla in the $\pm$ eigenstate of $\sigma_{x}$ corresponds to the map:
\be
\rho\otimes\rho\rightarrow\rho^{\prime}(\pm)=\frac{1}{4p(\pm)}\sum_{i,j}p_{i}p_{j}(|\phi_{i}\RR_{A}|\phi_{j}\RR_{B}\pm|\phi_{j}\RR_{A}|\phi_{i}\RR_{B})({_{A}\LL}\phi_{i}|{_{B}\LL}\phi_{j}|\pm{_{A}\LL}\phi_{j}|{_{B}\LL}\phi_{i}|).
\ee
The reduced states of the output state $\rho^{\prime}(\pm)$ are again identical but now equal to: $Tr_{A}[\rho^{\prime}(\pm)]=Tr_{B}[\rho^{\prime}(\pm)]=(\rho\pm \rho^2)/(2p(\pm))$.  Because $\rho^{\prime}(\pm)$ is an eigenstate of SWAP, subsequent measurements using the entanglement witness will reveal no further information about the state overlap $Tr[\rho^{2}]$.  Further, without knowledge of the input state it is not possible to unitarily disentangle the two systems.  However, a significant advantage is that the measurement consumes only two copies of the quantum state.  This should be compared to quantum state tomography on a $d$ dimensional system where $d^{2}-1$ measurements are needed to reconstruct the state and the same number of copies are consumed.  In the case of qubits, a measurement of purity using the entanglement witness is 50\% more efficient than performing state tomography.  For both techniques, the measurement statistics on $N$ identically prepared ensembles converge to the correct value of purity with an error that scales like $O(1/\sqrt{N})$.       

\begin{figure} [htbp]
\centerline{\epsfig{file=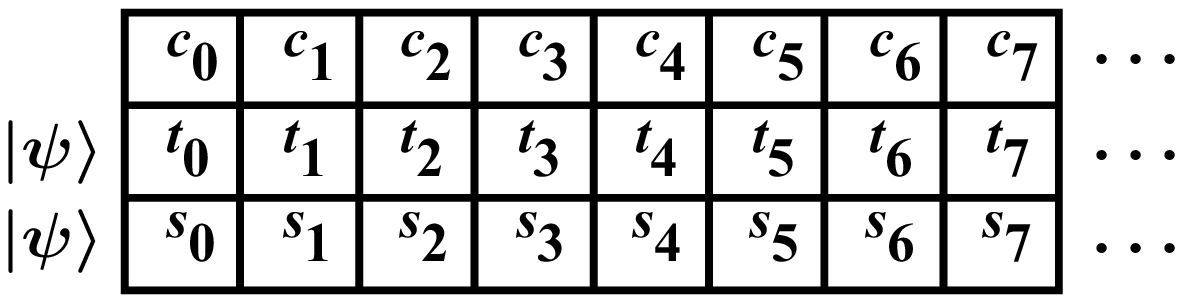, width=8.2cm}} 
\vspace*{13pt}
\fcaption{\label{fig:1}Stacked array of qubit registers as a platform to measure entanglement properties.  Two registers of $n$ spins each, denoted $t$ and $s$, are identically prepared in the state $|\psi\RR$ and interact with a third register of qubits $c$ prepared in $[1/\sqrt{2}(|0\RR+|1\RR)]^{\otimes n}$.  A sequence of gates implements the controlled-SWAP gate in parallel over the columns, with the qubits labelled $c_{j}$ acting as controls to the targets $t_{j},s_{j}$.  After the sequence, the $c$ register is measured in the $\sigma_{x}$ basis to extract information about the entanglement of $|\psi\RR$.   }
\end{figure}

In order to measure $Q$ for the state of a chain of spins, it is sufficient to measure the purity of each spin.  A schematic for implementing such a measurement in an efficient, parallel manner is shown in Fig.~\ref{fig:1}.  The setup includes a control register $c$ and two identically prepared target registers $t$ and $s$ all of length $n$.  It is assumed that  interactions between qubits within a register can be suppressed during the measurement process so that interactions take place only between triplets of qubits $\{c_{j},t_{j},s_{j}\}$, where $j$ is the site index.  The c-SWAP gate can be implemented in parallel over the columns of qubits with only nearest neighbor $\sigma_{z}\otimes\sigma_{z}$ type interactions between pairs $(c_{j},t_{j})$ and $(t_{j},s_{j})$ as we now show.

In deriving a pulse sequence for c-SWAP, we first point out that the SWAP gate, which is generated by the Heisenberg interaction $H_{exc}=g\vec{\sigma}\cdot\vec{\sigma}$, can be simulated by the Ising interaction, $H_{zz}=g\sigma_{z}\otimes\sigma_{z}$, by conjugation with single qubit gates as:
\be
\begin{array}{lll}
{\rm SWAP}(t,s)&=&e^{i\frac{\pi}{4}}e^{-i \frac{\pi}{4}(\sigma_{x}^{t}\otimes\sigma_{x}^{s}+\sigma_{y}^{t}\otimes\sigma_{y}^{s}+\sigma_{z}^{t}\otimes\sigma_{z}^{s})}\\
&=&e^{i\frac{\pi}{4}}e^{i\frac{\pi}{4}(\sigma_{y}^{t}+\sigma_{y}^{s})} e^{-i \frac{\pi}{4}\sigma_{z}^{t}\otimes\sigma_{z}^{s}} e^{-i\frac{\pi}{4}(\sigma_{y}^{t}+\sigma_{y}^{s})}e^{i\frac{\pi}{4}(\sigma_{x}^{t}+\sigma_{x}^{s})}\\
& & e^{-i\frac{\pi}{4} \sigma_{z}^{t}\otimes\sigma_{z}^{s}} e^{-i\frac{\pi}{4}(\sigma_{x}^{t}+\sigma_{x}^{s})}e^{-i \frac{\pi}{4} \sigma_{z}^{t}\otimes\sigma_{z}^{s}}.\;
\end{array}
\ee 
This simulation minimizes the time spent during two qubit interactions \cite{Glaser}.  To implement the c-SWAP gate, we replace $e^{-i \frac{\pi}{4} \sigma_{z}^{t}\otimes\sigma_{z}^{s}}$ in the above sequence with $e^{-i \frac{\pi}{4} |1\RR_{c}{_{c}\LL}1|\otimes\sigma_{z}^{t}\otimes\sigma_{z}^{s}}= e^{-i \frac{\pi}{8} \sigma_{z}^{t}\otimes\sigma_{z}^{s}}e^{i \frac{\pi}{8} \sigma_{z}^{c}\otimes\sigma_{z}^{t}\otimes\sigma_{z}^{s}}$.
The three body interaction can be simulated with pairwise interactions as,
\be
\begin{array}{lll}
e^{i \phi \sigma_{z}^{c}\otimes\sigma_{z}^{t}\otimes\sigma_{z}^{s}}&=&e^{-i\frac{\pi}{2}\sigma_{y}^{c}} e^{-i\frac{\pi}{4}\sigma_{y}^{t}} e^{-i\frac{\pi}{4}\sigma_{x}^{t}} e^{-i \frac{\pi}{4} \sigma_{z}^{c}\otimes\sigma_{z}^{t}}e^{i\frac{\pi}{4}\sigma_{y}^{c}} e^{i\frac{\pi}{4}\sigma_{x}^{t}}e^{-i \phi \sigma_{z}^{t}\otimes\sigma_{z}^{s}}\\
& & e^{-i\frac{\pi}{4}\sigma_{y}^{c}} e^{-i\frac{\pi}{4}\sigma_{x}^{t}}e^{i \frac{\pi}{4} \sigma_{z}^{c}\otimes\sigma_{z}^{t}} e^{i\frac{\pi}{2}\sigma_{y}^{c}} e^{i\frac{\pi}{4}\sigma_{x}^{t}} e^{i\frac{\pi}{4}\sigma_{y}^{t}}.\;
\end{array}
\label{threebody}
\ee
The total sequence for the c-SWAP is therefore:
\be
\begin{array}{lll}
{\rm c-SWAP}(c,t,s)&=&e^{-i\frac{\pi}{8}\sigma_{z}^{c}}
e^{i\frac{\pi}{4}(\sigma_{y}^{t}+\sigma_{y}^{s})} e^{-i \frac{\pi}{8} \sigma_{z}^{t}\otimes\sigma_{z}^{s}}e^{i \frac{\pi}{8} \sigma_{z}^{c}\otimes\sigma_{z}^{t}\otimes\sigma_{z}^{s}} e^{-i\frac{\pi}{4}(\sigma_{y}^{t}+\sigma_{y}^{s})}e^{i\frac{\pi}{4}(\sigma_{x}^{t}+\sigma_{x}^{s})}\\
& & e^{-i \frac{\pi}{8} \sigma_{z}^{t}\otimes\sigma_{z}^{s}}e^{i \frac{\pi}{8} \sigma_{z}^{c}\otimes\sigma_{z}^{t}\otimes\sigma_{z}^{s}}e^{-i\frac{\pi}{4}(\sigma_{x}^{t}+\sigma_{x}^{s})} e^{-i \frac{\pi}{8} \sigma_{z}^{t}\otimes\sigma_{z}^{s}}e^{i \frac{\pi}{8} \sigma_{z}^{c}\otimes\sigma_{z}^{t}\otimes\sigma_{z}^{s}}.\;
\end{array}
\ee

For two qubit gates generated by pairwise Hamiltonians, $H_{zz}$, with $g>0$, the total interaction time needed to implement the c-SWAP gate using this sequence is $T_{{\rm c-SWAP}}=27\pi/(4g)$.  If the sign of $g$ is tunable to $g<0$, then the interaction time is reduced to $T_{{\rm c-SWAP}}=9\pi/(4|g|)$.   After performing  c-SWAP gates in parallel, a measurement of the qubits in register $c$ in the $\sigma_{x}$ basis will yield a mean number of $1's$ equal to: $n\langle p(+)\rangle=nQ(|\psi\RR)/4$. 

It will be important to verify that the state of a system $s$ is a pure state.  If it can be assumed that the two systems $s$ and $t$ are identically prepared, each in state $\rho$, then one can measure the purity of the entire state by a simple adaptation of the above procedure.  If the $n$ bit control register $c$ is prepared in the initial state $|\phi(0)\RR=1/\sqrt{2}(|0\RR+|1\RR)_{0}\otimes |0\ldots0\RR$, then a sequence of gates $U=\bigotimes_{k=n-1}^{1}{\rm CNOT}(k-1,k)$ will produce the n-partite GHZ state in $c$.  Now the c-SWAP gate over the entire states $s$ and $t$ can be applied bit-wise over the triplets of qubits  $(c_{j},t_{j},s_{j})$.  After this the control register is disentangled by operating with $U^{-1}$ and the qubit $c_{0}$ is measured in the $\sigma_{x}$ basis to measure the purity of the state $\rho$.  Using entangled states of the control system allows a measurement to infer the purity of any subset of qubits in the state of $s$ simply by preparing an entangled state of the corresponding control qubits and applying the c-SWAP gate over that subset.  In this way, states with sub-global entanglement can be distinguished from globally entangled states not separable over any subsystem.  After measuring the purity of a given subset of qubits, the constituent qubits in the two registers $s$ and $t$ will become entangled with each other.  In order to disentangle them and reuse them, the state of the qubits in the measured set of the $t$ register will need to be depolarized and reset to $|0\RR$.  In general, unless there exists prior knowledge of the distribution of entanglement across the system, this technique for distinguishing different partitions of entanglement is inefficient because the number of partitions of the $n$ qubits is $2^{n}$ .       

A testbed for implementing this multi-qubit entanglement measurement is an ensemble of trapped neutral atoms in a three dimensional optical lattice.  This system has the advantage of inherent parallelism for quantum processing over identically prepared registers of atomic qubits.   There are several proposals for QIP in optical lattices with varying demands on addressability of the atomic qubits.  Because $Q$ is a quantity related to an average signal over the control register, some constraints on addressability within the system can be relaxed.  If it can be safely assumed that state of the register is pure, then by using the three neighboring lattices as explained above, one does not need to resolve qubits within each register but only between registers $c$, $t$, and $s$.  Therefore, the following protocol should be applicable both to implementations using a polarization gradient lattice along one dimension where the inter-atomic spacing is not directly optically resolvable \cite{Nonaddress} and schemes where the atoms are trapped along that dimension by long wavelength light of uniform polarization \cite{Rydberg, Charron}.  Atoms can be trapped along the other two directions by two pairs of counter-propagating linearly polarized beams that are distinct in frequency so that they do not interfere with each other.  Such a configuration produces planes of identical one dimensional registers with planar separation along the $\hat{x}$ direction.  In order to measure the multi-qubit entanglement in a given register, it is necessary to have a mechanism to interact neighboring planes of qubits, and have an inter-plane separation much larger than an optical wavelength so that the planes can be addressed.  A possible solution is to use a bi-periodic potential to trap along $\hat{x}$ as is suggested in \cite{Charron}.  There it is proposed to trap ${^{87}{\rm Rb}}$ atoms along one dimension with two pairs of counter-propagating linearly polarized laser beams from the fundamental and first harmonic of a ${\rm CO}_{2}$ laser.  By adjusting the relative intensities of the two colors of light, pairs of atoms can be made to interact along that dimension.  

\begin{figure} [htbp]
\centerline{\epsfig{file=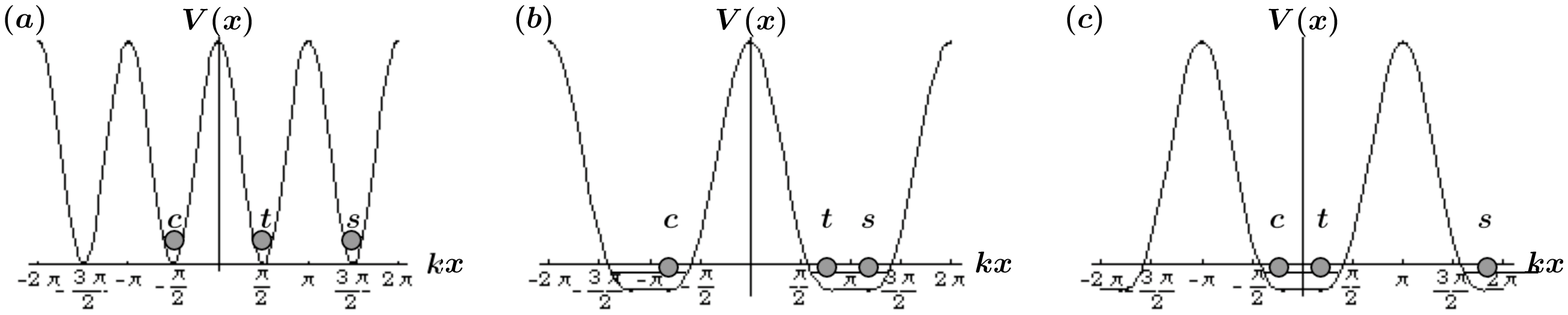, width=11.2cm}} 
\vspace*{13pt}
\fcaption{\label{fig:2}Cross section of 3D optical lattice shown as a platform to implement a measurement of entanglement.  The trapping potential along $\hat{x}$ results from an incoherent sum of two light fields and can be tuned to join planes of trapped atoms, ordered $(c,t,s)$, together pairwise.  (a)  Initial configuration with $V_{1}=0$, $\delta=0$.  (b) $V_{1}/V_{0}=0.3$, $\delta=0$.  (c) $V_{0}/V_{1}=0.3$, $\delta=\pi/2$.}
\end{figure}

Consider three planes of atoms, ordered $(c,t,s)$, localized along $\hat{x}$ at the minima of an external potential $V(x)=V_{0}\cos^{2}(kx)$ with inter-planar separation $\delta x=\pi/k$.  The planes could each constitute a single quantum register, or an array of 1D registers.  Introducing a second light field, phase locked with respect to the first by a phase $\delta$,  and at half the frequency creates a bi-periodic potential: $V(x)=V_{0}\cos^{2}(kx)+V_{1}\cos^{2}(kx/2+\delta)$.  If the second field is turned on adiabatically with respect to the oscillation period of the wells created by the first potential, then the motional states of the atoms will follow the ground state of the new potential.  As shown in Fig.~\ref{fig:2}, by adjusting the relative intensity and phase of the two fields, atoms in planes $(t,s)$ or $(c,t)$ can be joined pairwise at the two minima of a double well potential.  The double well barrier height can be adjusted to allow substantial overlap in the motional wavefunctions of each atom in neighboring planes.  If the quantum information is encoded in the proper ground state electronic or motional levels, a controlled ground-ground state collision can be engineered to generate evolution by an Ising type interaction \cite{Note}.  Provided the fundamental trapping frequency is chosen so that $\delta x>\pi/k_{res}$, where $ck_{res}/(2\pi)$ is the resonant atomic frequency, the 3 planes of atoms will be addressable.  Because the potential period actually depends on the difference wave-vector of the counter-propagating beams, addressability can also be mandated by using beams that are not parallel, but rather intersect at an angle $\theta$ so that $\delta x$ is increased by a factor of $1/\sin(\theta)$.  The c-SWAP gates can be implemented in parallel and the values of the qubits in the control register can be read out at once using a spatially broad laser on a resonant cycling transition.  The observed resonant fluorescent signal will be proportional to the average purity. 

\section{Conclusions}
We have shown that a previously introduced entanglement measure on pure states of an arbitrary number of qubits is simply expressed as a linear function of the average purity of the qubits.  This quantity can be observed in a straightforward way in the laboratory either by performing state tomography on each system qubit or by a more efficient technique using bitwise interactions with a second identically prepared register.  The latter protocol is particularly well suited for systems for QIP with massive inherent parallelism such optical lattices.  It is hoped that other, directly observable, measures will be found that more sharply distinguish the partitioning of shared entanglement over multi-partite systems.

\nonumsection{Acknowledgements}
\noindent
I thank Jamie E. Williams, Manny E. Knill and Carlton Caves for helpful discussions.  I also appreciate correspondence with Todd Brun who pointed out an error in the first draft of the manuscript and the comments of anonymous referees.  This work was supported in part by ARDA/NSA.

\nonumsection{References}


\begin{thebibliography}{000}
\bibitem{Note} The particular implementation will determine the proper encoding.  For instance, to interact atoms via controlled ground state collisions, the magnetic moments of the atoms may need to be rotated to the new quantization axis $\hat{x}$ before the c-SWAP sequence.
\end{thebibliography}
\end{document}